\documentclass[useAMS,usenatbib,usegraphicx]{mn2e}
\usepackage{graphics}
\usepackage{graphicx}
\usepackage{psfig}
\usepackage{psfrag}
\usepackage{amsmath}
\usepackage{amssymb}
\usepackage{ulem}
\usepackage{ctable}
\usepackage{txfonts}
\usepackage{enumerate}
\usepackage{url}
\usepackage[colorlinks=true,linkcolor=blue,urlcolor=blue,citecolor=blue]{hyperref}
\usepackage{color}

\newcommand{\plk}{{\it Planck~}}

\newcommand{\comm}{}
\newcommand{\newcomm}{}
\newcommand{\fincomm}{}

\defcitealias{Brown11}{BR11}
\defcitealias{Planck13}{P13}
\defcitealias{Akamatsu13}{A13}

\title[SZE measurement of the Coma relic]
{Evidence for a pressure discontinuity at the position of the Coma relic from \plk Sunyaev-Zel'dovich effect data}

\author[J. Erler et al.]{J. Erler\thanks{E-mail: jens@astro.uni-bonn.de}, 
~K. Basu\thanks{E-mail: kbasu@astro.uni-bonn.de}, 
~M. Trasatti, ~U. Klein, ~F. Bertoldi.\\
Argelander Institut f\"ur Astronomie, Universit\"at Bonn, Auf dem H\"ugel 71, 53121 Bonn, Germany
}

\begin{document}

\date{Accepted; Received; in original form}

\pagerange{\pageref{firstpage}--\pageref{lastpage}}\pubyear{2014}

\maketitle
\label{firstpage}

\begin{abstract}
\noindent 
Radio relics are Mpc-scale diffuse synchrotron sources found in galaxy cluster outskirts. 
They are believed to be associated with large-scale shocks propagating through the intra-cluster medium, although the connection between radio relics and the cluster merger shocks is not yet proven conclusively. 
We present {\newcomm a first tentative detection} of a pressure jump in the well-known relic of the Coma cluster through Sunyaev-Zel'dovich (SZ) effect imaging.
The SZE data are extracted from the first public all-sky data release of \plk and we use high-frequency radio data at $2.3 \, {\rm GHz}$ to constrain the shock-front geometry. 
The SZE data provide evidence for a pressure discontinuity, {\comm consistent with the relic position, without requiring any additional prior on the shock-front location}.  
The derived Mach number ${\cal M} = 2.9^{+0.8}_{-0.6}$ is consistent with X-ray and radio results. 
A high-pressure ``filament'' without any pressure discontinuity is disfavoured by X-ray measurements and a ``sub-cluster'' model based on the infalling group NGC 4839 can be ruled out considering the published mass estimates for this group. 
{\newcomm These results signify a first attempt towards directly measuring the pressure discontinuity for a radio relic} and the first SZ-detected shock feature observed near the virial radius of a galaxy cluster.
\end{abstract}


\section{Introduction}

Radio relics in galaxy clusters are elongated, diffuse synchrotron sources up to an Mpc in length, which are not directly associated with the AGN activity in galaxies. 
They are typically found near cluster virial radii and are characterized by a steep radio spectrum ($\alpha > 1$, where $S_{\nu} \propto \nu^{-\alpha}$) and a high degree of polarization ($20\% - 30\%$ at $1.4 \, {\rm GHz}$). About 50 radio relics are known, mostly in low redshift clusters (see review by Feretti et al. \citeyear{Feretti12}). One of the best-known examples is the radio relic in the Coma cluster (Giovannini, Feretti \& Andernach \citeyear{Giovannini85}).  
It is generally assumed that radio relics mark the location of shock fronts resulting from cluster merger events. The shock waves can amplify particle energies through the \mbox{Fermi-I} process as well as amplify the magnetic fields, resulting in the strong synchrotron emissions seen as radio relics (e.g. En{\ss}lin et al. \citeyear{Ensslin98}; Miniati et al. \citeyear{Miniati01}; Hong et al. \citeyear{Hong14}; see also Guo, Sironi \& Narayan \citeyear{Guo14}). However, there are several shortcomings to this accepted picture and the relevant details are not well understood. 
Merger shocks in the cluster periphery are expected to have low Mach numbers (${\cal M} \sim 2-4$), so the particle acceleration efficiency would be very small. Phenomenologically, a relic is not always detected at cluster shock fronts and also most of the known relics do not have a detected shock feature. 

Direct detections of shock discontinuities at radio relic positions have so far only been obtained through X-ray observations. This is inherently challenging, as the peripheral location of radio relics in clusters means that their X-ray {\comm emission} can be dwarfed by the high background noise. 
It is generally easier to detect an X-ray surface brightness discontinuity at radio relic positions, but the corresponding density jumps alone can not distinguish between shocks and cold fronts (e.g. Markevitch \& Vikhlinin \citeyear{Markevitch07}) and modelling density discontinuities is more prone to projection biases than temperature. 
For the well-known Coma relic, tentative determination of a temperature discontinuity has been reported from the analysis of {\it Suzaku} (Akamatsu et al. \citeyear{Akamatsu13}, hereafter \citetalias{Akamatsu13}) and {\it XMM-Newton} data (Ogrean \& Br{\"u}ggen \citeyear{Ogrean13}), whereas another analysis of the {\it Suzaku} data failed to detect a temperature drop (Simionescu et al. \citeyear{Simionescu13}). 
 
An independent method for confirming the presence of shock waves at radio relic positions would be to detect the pressure discontinuity directly through the thermal Sunyaev-Zel'dovich effect (tSZE; or simply SZE) (Sunyaev \& Zel’dovich \citeyear{Sunyaev72}). 
Recent instrumental advances have opened up the SZE as a promising tool to study cluster astrophysics (Ferrari et al. \citeyear{Ferrari11}; Korngut et al. \citeyear{Korngut11}; Planck Collaboration \citeyear{Planck13}, hereafter \citetalias{Planck13}). 
Yet no direct determination of the shock front at radio relics with the SZE effect has been published. 
In this Letter, we report a first {\comm tentative} detection of a pressure discontinuity at the position of a radio relic from the analysis of SZE data for Coma cluster. The maps used to extract SZ-Comptonization images were taken from the first public data release of {\it Planck}. In addition, we use new $2.3 \, {\rm GHz}$ radio {\comm synthesis} imaging data to constrain the position and width of the shock front. 
We assume a flat $\Lambda$ cold dark matter Cosmology with $h = 70$, $\Omega_\Lambda = 0.7$ and $\Omega_\mathrm{M} = 0.3$. We adopt $z=0.0231$ from Struble \& Rood \citeyearpar{Struble99}, which implies $28 \, {\rm kpc} \, {\rm arcmin^{-1}}$ at Coma's distance and use $R_{500} = 47 \pm 1 \, {\rm arcmin}$ (\citetalias{Planck13}).

\vspace*{-4mm}
\section{Data analysis}
\label{sec:data}
\subsubsection*{Radio data:}
\vspace*{-1mm}
The radio relic in the Coma cluster was observed at $2.3 \, {\rm GHz}$ ($S$ band) with the Westerbork Synthesis Radio Telescope (WSRT) and at $2.64 \, {\rm GHz}$ with the Effelsberg-$100 \, {\rm m}$ radio telescope. Here, we present a brief outline of the radio data analysis and imaging method. More details will be given in a future publication (Trasatti et al., in prep.). 

\begin{figure*}
\centering
\includegraphics[width=0.97\columnwidth]{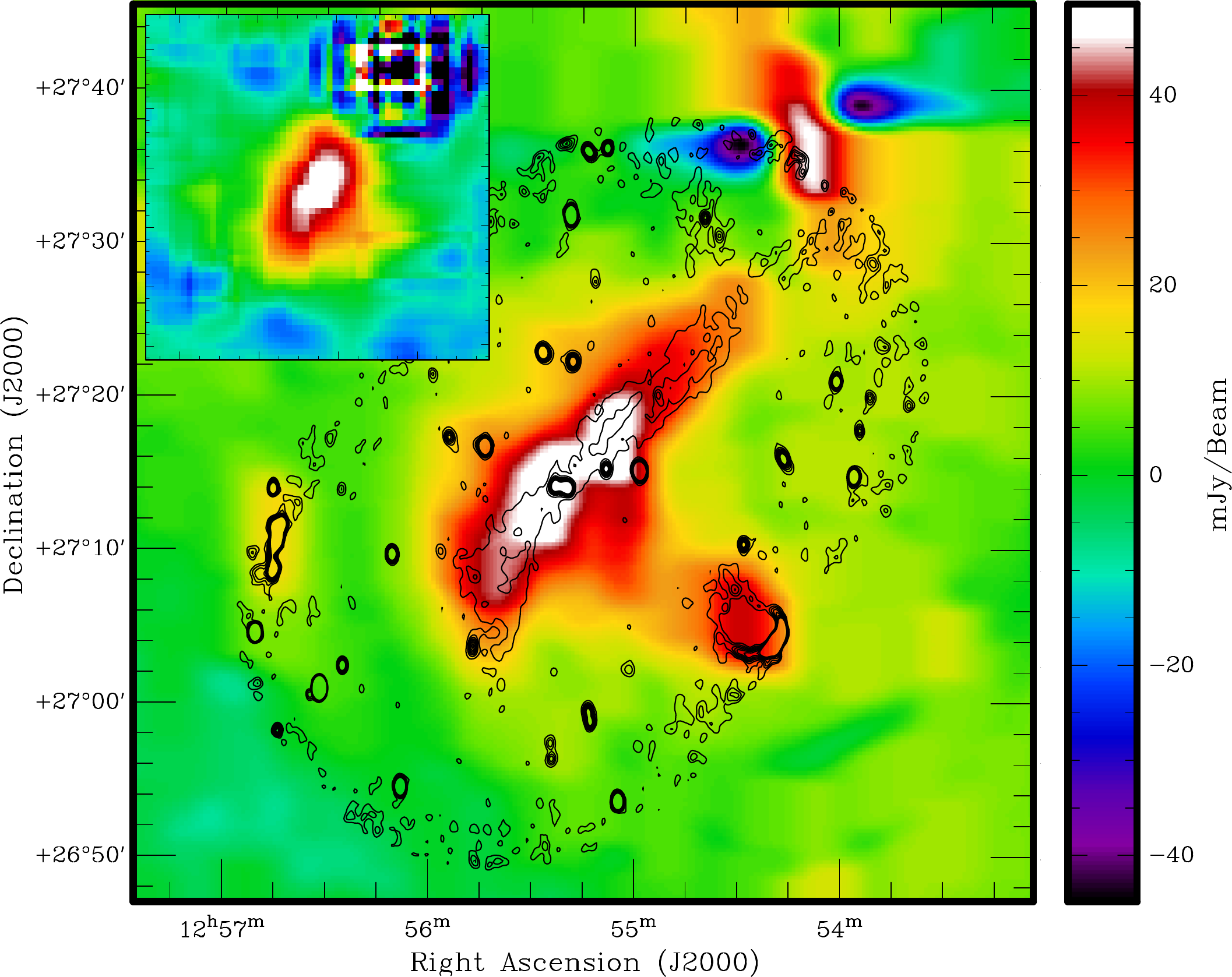}
\hspace*{12mm}
\includegraphics[width=0.97\columnwidth]{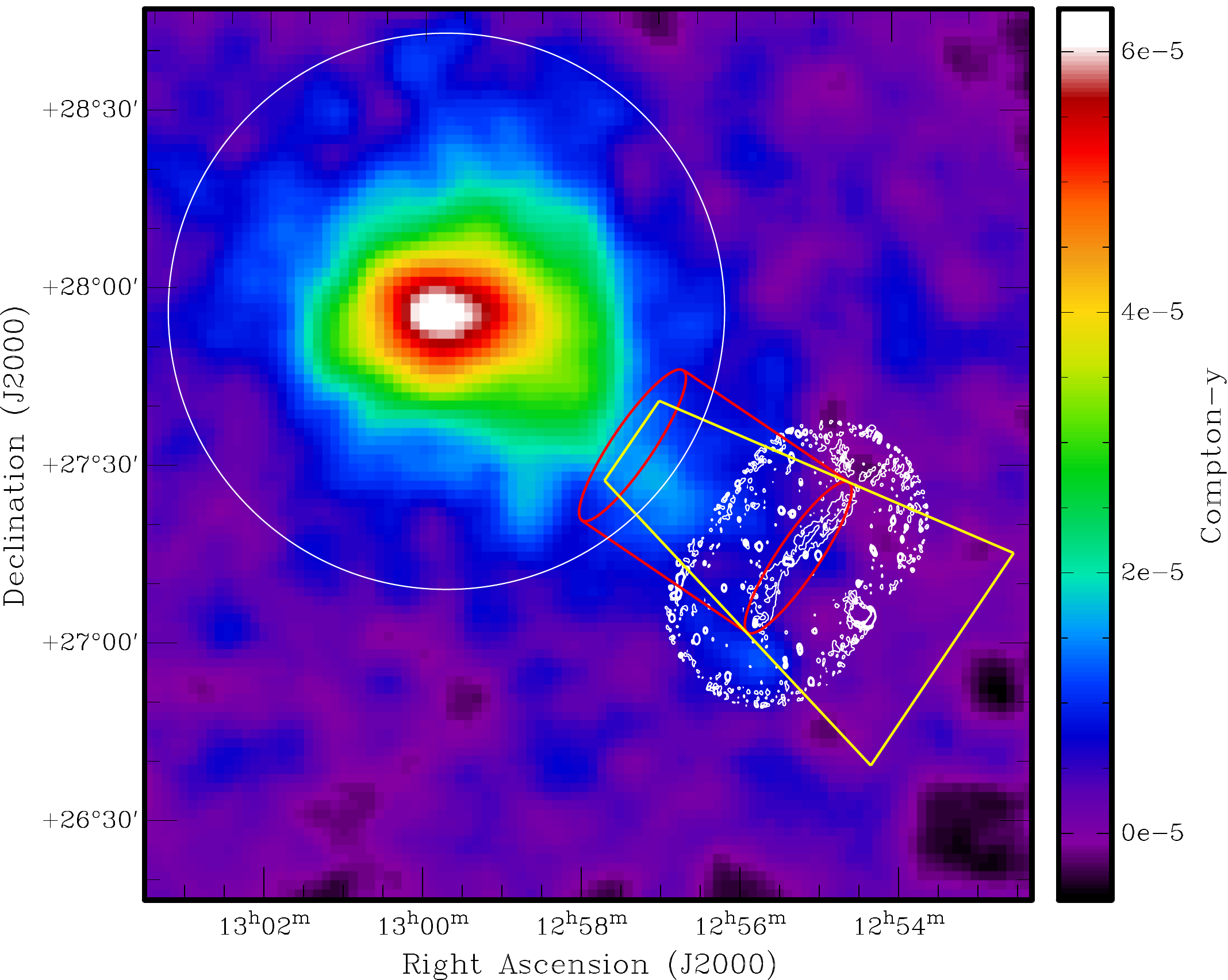}
\caption{Radio and Sunyaev-Zel'dovich (SZ) effect imaging maps for the Coma relic position. {\bf Left-hand panel:} low-frequency WSRT radio synthesis imaging at $350 \, {\rm MHz}$ in colour (from \citetalias{Brown11}), with overlaid high-frequency image (at $2.3 \, {\rm GHz}$) of the radio relic from our data in black contours. The low-frequency map has been filtered to remove point source contributions. The increased noise near the edge of the high-frequency radio map is due to primary beam corrections. {\comm The inset shows the same field around the Coma relic at $9.5 \, {\rm arcmin}$ resolution and was created by subtracting an $1.4 \, {\rm GHz}$ NVSS image from a new $L$-band Effelsberg image.} 
{\bf Right-hand panel:} the Comptonization map derived from the linear combination of \plk data. 
The thin white circle denotes the $R_{500}$ of Coma. The white contours, which are identical to the black ones in the left-hand panel, show the position of the relic, which is located at roughly $1.6\times R_{500}$. The red cylinder illustrates our assumed geometry of the {\comm 3D pressure tube}, consistent with the relic length, while the yellow wedge marks the {\comm 2D} region from which the profile used during our analysis is extracted.}
\label{fig:maps}
\end{figure*}

The WSRT field of view at this frequency is $\sim$$22 \, {\rm arcmin}$, mainly limited by the effect of the primary beam attenuation. Hence, the observations were carried out in a mosaic mode, with three different pointing centres and $4{\rm h}$ exposure time per pointing.
Primary beam correction was accounted for, due to which the rms noise increases from $0.04 \, {\rm mJy/beam}$ at the map centre to $0.12 \, {\rm mJy/beam}$ near map edges. 
The synthesized WSRT beam is roughly $33 \, {\rm arcsec} \times 13 \, {\rm arcsec}$. The minimum baseline for the WSRT configuration used is $36 \, {\rm m}$, which makes the imaging insensitive to structures larger than $\sim$$12.5 \, {\rm arcmin}$. Therefore, the WSRT observation were complemented with data from the Effelsberg-$100 \, {\rm m}$ telescope at $2.64  \, {\rm GHz}$. The total Effelsberg field size is $3^\circ \times3^\circ$ with an rms noise of $\sim$$2.3 \, {\rm mJy/beam}$ and is thus effective in providing the missing short-spacing information in the WSRT image. The two data sets were combined in the Fourier domain with the {\small IMERG} task in {\small AIPS}.  
The image is shown in Fig. \ref{fig:maps} (left-hand panel) with black contours, on a $325 \, {\rm MHz}$ WSRT radio image published by Brown \& Rudnick (\citeyear{Brown11}, hereafter \citetalias{Brown11}).
The radio data are primarily used to constrain the width of the shock front in our modelling (see Section \ref{sec:model}). High-frequency radio emission in a radio relic should mark the location of freshly accelerated electrons closest to the shock front, so in contrast to the available low-frequency data it would be a cleaner indicator of the shock front geometry.  

A critical issue for our modelling is the total length of the Coma relic, as the extremely bright background radio source Coma A (3C277.3) fully obscures the northern part of the relic in low-frequency data (e.g. \citetalias{Brown11}). In Fig. \ref{fig:maps} (left, colour image), the residual artefacts after subtracting the bright Coma A source can be seen north of the relic. 
It could be argued that the relic emission extends right through the position of Coma A (\citetalias{Brown11}). However, no clear indication of such an extension is seen from our high-resolution contours. As further proof, we construct a low-resolution ($9.4 \, {\rm arcmin} \times 9.4 \, {\rm arcmin}$) image of the Coma relic at $1.4 \, {\rm GHz}$ by subtracting an NVSS image (Condon et al.\citeyear{Condon98}) from a new $L$-band Effelsberg image (Fig. \ref{fig:maps} left, {\comm inset}), which also does not show the relic emission extending up to the masked Coma A position. 
The dimensions of the radio relic used in our modelling are therefore $728\pm14 \, {\rm kpc}$ ($26 \pm 0.5  \, {\rm arcmin}$) in the tangential direction and the radial thickness varies between $50 \ {\rm and} \ 125 \, {\rm kpc}$. 

\vspace*{-2mm}
\subsubsection*{Sunyaev-Zel'dovich effect (SZE) data:}
\vspace*{-1mm}
{\comm The thermal SZE is measured by the Comptonization parameter $y$, which is proportional to the line-of-sight integral of the electron pressure: 
$y = (\sigma_{\rm T}/m_{\rm e}c^2) \int P_{\rm e} \, {\rm d}\ell$.}
We construct a $30^{\circ} \times 30^{\circ}$ $y$-map of Coma from the \plk data by forming linear combinations of maps (the ILC method; Bennett et al. \citeyear{Bennett03}; Remazeilles et al. \citeyear{Remazeilles11}), using all six frequency bands of the \plk High Frequency Instrument (HFI), taken from the nominal 15-month survey. 

A Gaussian filter was applied to smooth all maps to a common resolution of $10 \, {\rm arcmin}$, corresponding to the \plk beam at $100 \, {\rm GHz}$. The final Compton-$y$ map is the weighted sum of all six maps: $y = \sum_i w_i T_i / T_{\mathrm{CMB}}$. Here, $T_i$ are the individual channel maps, each weighted with an ILC-coefficient $w_i$. 
The coefficients are chosen to minimize the variance of the reconstructed Compton-$y$ map while fulfilling two constrains: (1) {\comm eliminate} the primary CMB temperature anisotropies and (2) preserve the temperature fluctuations introduced by the SZE. The produced map may contain a signal offset, since the variance of the map stays {\comm unaffected} while adding a constant. To correct for this offset, we subtracted the map mean which was calculated while excluding a circular region with a radius of $5 \times R_{500}$ centred around the cluster.
The resulting Compton-$y$ map of the Coma cluster is shown in Fig. \ref{fig:maps} (right). Its noise level on pixel scale is $\sigma_{\mathrm{pix}} = 2.2 \times 10^{-6}$. 

Our implementation of the ILC method used to obtain a Coma $y$-map does not include any foreground contaminants, e.g. galactic dust and synchrotron emission. The $30^{\circ} \times 30^{\circ}$ field around Coma is relatively dust free and we find excellent agreement with the published \plk results on Coma (\citetalias{Planck13}) on the profile errors and overall noise properties, even though the \plk team implemented more sophisticated multi component ILC methods (MILCA; Hurier, Mac{\'{\i}}as-P{\'e}rez \& Hildebrandt \citeyear{Hurier13}). We assumed the radio contamination from Galactic synchrotron and background point sources to be negligible at HFI frequencies. We also produced a $y$-map including the LFI $70 \, {\rm GHz}$ channel, but found no significant improvement of the signal-to-noise ratio for Coma, despite a degradation of resolution to $13 \, {\rm arcmin}$. Therefore, our results are based on the six HFI maps.


\vspace*{-4mm}
\section{Models}
\label{sec:model}

\begin{figure}
\includegraphics[height=0.94\columnwidth, angle=-90]{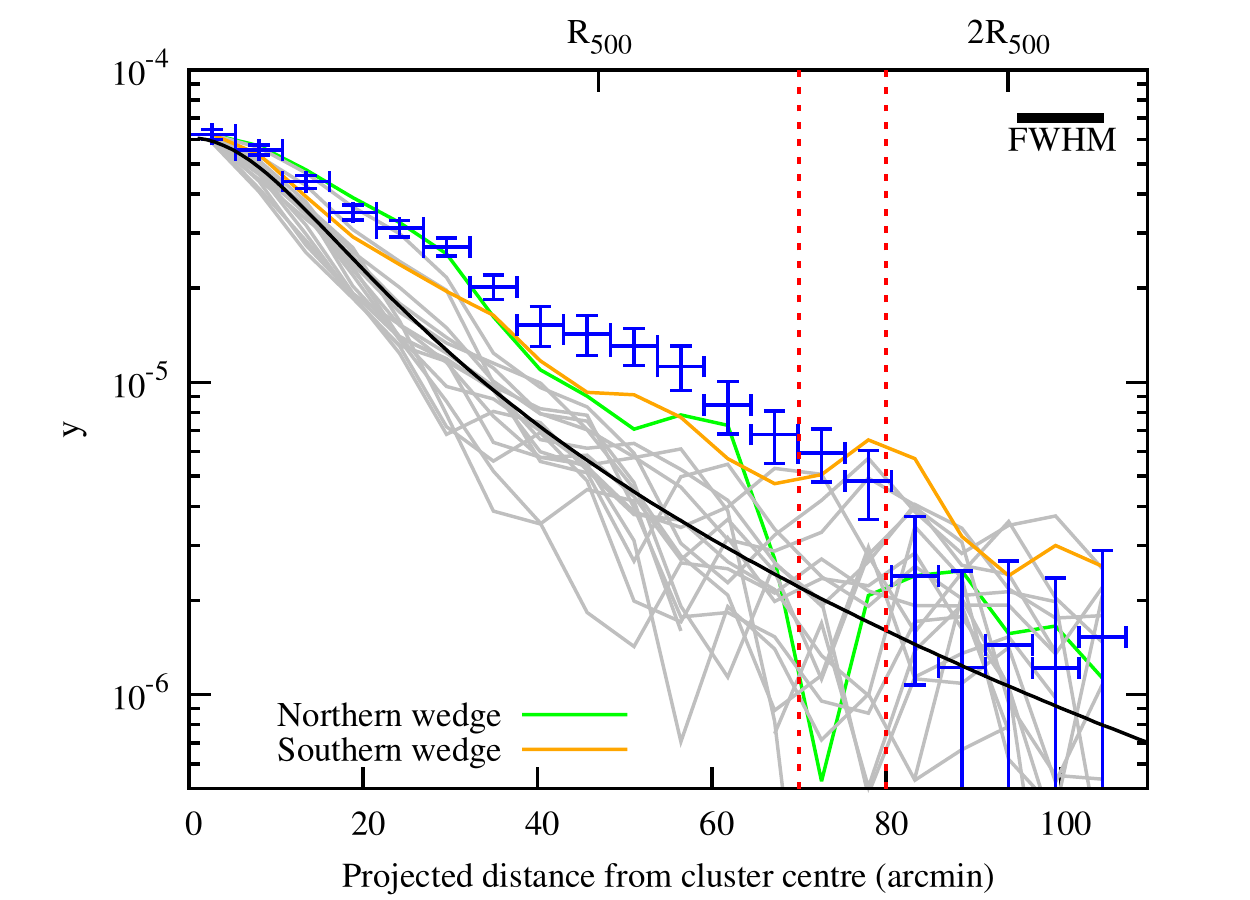}
\caption{SZE profiles of the Coma cluster. The blue data points with errors mark the radial profile along the relic direction, extracted from a wedge-shaped region of opening angle 20$^{\circ}$. The thin grey lines mark similar radial profiles in all the other directions. The two neighbouring wedge profiles below and above the relic-wedge are  marked in colour. The black continuous line is the GNFW model fit for the average radial profile, excluding the wedge that contains the radio relic. The two vertical dotted lines mark the prior boundaries on the radio relic position derived from $2.3 \, {\rm GHz}$ radio data. {\comm The increase of $y$ in the southern wedge close to the distance of the relic is likely due to a background galaxy cluster located near ${\rm (RA, Dec)}=(12^{\rm h}55^{\rm m}42^{\rm s}, \ +26^{\circ}57'11'')$}.}
\label{fig:allcones}
\end{figure}

To assess whether a pressure discontinuity can be identified at the relic position, we analyse the radial $y$-profiles derived from the  \plk data. Fig. \ref{fig:allcones} shows the result of a radial profile extraction along 18 identical {\comm wedges} of angle $20^{\circ}$ from the Coma cluster centre at ${\rm (RA, Dec.)}=(12^{\rm h}59^{\rm m}47^{\rm s}, \ +27^{\circ}55'53'')$. 
The blue data points are the projected pressure along the direction of the relic, whose errors are computed from the covariance of 1000 similar wedge-shaped regions in the $30^{\circ} \times 30^{\circ}$ $y$-map. The broken solid lines mark the $y$-profiles in the other directions, with the two neighbouring wedges marked separately in colour. 
Clearly, there is an excess of pressure in the region between $R_{500}$ and $2 \, R_{500}$ of the cluster in the relic direction, roughly terminating at the edge of the relic (the relic width, as measured from the radio data, is indicated by the two vertical dotted lines). Within $R_{500}$, the excess pressure comes from a general elongation of the Coma cluster in this direction, as also noted by the \citetalias{Planck13} analysis. Beyond the relic position, the data are generally consistent with noise for all wedges. We can see an indicative jump of the projected pressure where the radio relic ends, but its significance is low. To quantify this small effect and deproject the underlying pressure discontinuity, we explore two classes of models. 

\newpage 

\noindent (1) {\it Pressure-tube model:} the $y$-excess in the SW direction of Coma resembles a high-pressure filament, either caused by the passage of a shock front, or a pre-existing filament on which the radio relic resides. To model this, we assume a cylindrical high-pressure ``tube'' geometry in the plane of the sky whose diameter is determined by the length of the relic (Fig. \ref{fig:maps} right). 
The region within the cluster's $R_{500}$ are excluded from {\newcomm all pressure excess model fits}.  
Inside this {\comm high-pressure} tube, we model the pressure with a power-law function {\newcomm that depends only on the radial distance} from the Coma centre: $P_{\rm tube}(r) = P_{\rm s} (r / r_{\rm s})^{\gamma}$, where $r$ is the radial distance, $r_{\rm s}$ is a constant and $P_{\rm s}$ and $\gamma$ are the model parameters. This pressure tube is placed over a spherically symmetric generalized Navarro-Frenk-White (GNFW) pressure model (Nagai, Kravtsov \& Vikhlinin \citeyear{Nagai07}; Arnaud et al. \citeyear{Arnaud10}) for the Coma cluster, which is constrained from the SZ-data excluding the radio relic wedge. We let the pressure excess terminate at the radio relic position (or any arbitrary position determined by the SZE data alone), beyond which the SZE signal is contributed only from the main cluster's GNFW pressure. This we term as the ``shock model''. 
In a variation to this scheme, we also allow the power-law pressure excess to continue uninterrupted across the radio relic, which we call the ``filament'' model.

\smallskip 

\noindent (2) {\it Sub-cluster model:} as an alternative to the pressure tube scenario, we construct a model where the pressure excess in the relic-wedge is caused purely by the presence of the infalling galaxy group NGC 4839, which is seen as a ``lump'' near the inner boundary of the relic-wedge (Fig. \ref{fig:maps} right). This galaxy group was first identified as a substructure of the Coma cluster by Mellier et al. \citeyearpar{Mellier88} in the optical. 
Observations with {\it XMM-Newton} (Neumann et al. \citeyear{Neumann01}) revealed a complex temperature structure around the group, which suggests the group to be at its first infall into Coma.  
We model the group as a second GNFW component with 1/10 the mass of the Coma cluster, which is the upper mass limit derived from a {\comm dynamical analysis} (Colless \& Dunn \citeyear{Colless96}). Alternatively, we let the mass be a free parameter, i.e. we fit the pressure amplitude while keeping the shape constant.


\vspace*{-6mm}
\section{Results}
\label{sec:results}

\begin{figure*}
\centering
\vspace*{-3mm}
\includegraphics[width=10.35cm, angle=270]{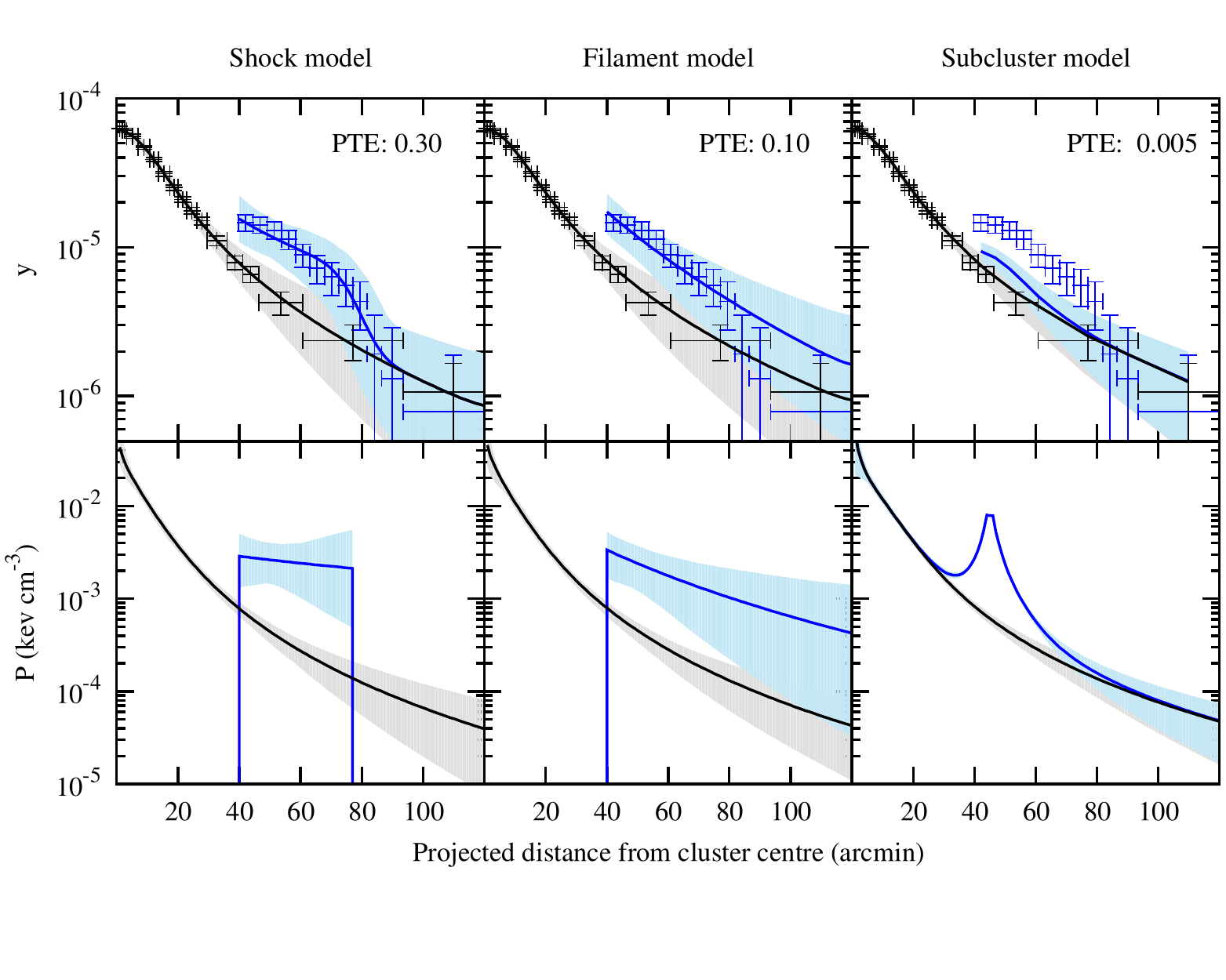}
\vspace*{-6mm}
\caption{Models for the pressure excess in the direction of the radio relic (lower panels) and the resulting fits to the observed $y$-profile (upper panels). {\comm Blue colours refer to the high pressure region in the SW of the Coma cluster, black and grey represent the mean cluster profile}. 
{\bf Left-hand panels}, the ``shock model'': the pressure excess terminates at the radio relic position, beyond which the pressure is described by the mean GNFW profile of the Coma cluster.
{\bf Middle panels}, the ``filament model'': here, we assume no pressure discontinuity at the relic position and that the power-law profile extends throughout the pressure tube (whose diameter is still determined by the radio relic). 
{\bf Right-hand panels}, the ``sub-cluster model'': here, we fit the infalling group NGC 4839 with an additional GNFW pressure model with fixed amplitude and shape parameters, such that it corresponds to a halo with 10\% of Coma mass.  
In all models, we exclude the inner $\sim 40 \, {\rm arcmin}$ region in the relic-wedge for fits to the excess pressure. The shaded regions indicate the envelope derived from the 68.3\% of models with the highest likelihood.}
\label{fig:results}
\end{figure*}

The main cluster's GNFW model and the second model for the high pressure region in the direction of the relic are fitted simultaneously through a Bayesian approach using Markov Chain Monte Carlo (MCMC) sampling. We radially bin the data inside and outside the high-pressure {\comm wedge} separately. Data inside the 20$^{\circ}$ relic-wedge are divided into 12 bins ($r_{min}=0.85\, R_{500}$, $r_{max}=2.7\, R_{500}$), whereas the global cluster profile is extracted from 16 circular regions excluding the relic-wedge ($r_{min}=0$, $r_{max}=2.7\, R_{500}$). The bin-to-bin noise covariance is extracted from  $1000$ similarly shaped regions at random positions across the $30^{\circ} \times 30^{\circ}$ $y$-map, while excluding the inner $5 \, R_{500}$. 
The 3D pressure models are projected in the plane of the sky and convolved with the \plk beam. We account for the mismatch between the wedge and the high-pressure cylinder{\newcomm, as well as for the impact of {\it Planck's} beam on to the assumed shock geometry.} 
The mean GNFW model fit is shown in Fig. \ref{fig:allcones} as a solid black curve. It provides an excellent fit to the radial data {\fincomm with the following parameters: 
\hbox{$(P_0, c_{500}, \gamma, \alpha, \beta) = (4.03^{+0.65}_{-0.56}, 2.03^{+0.61}_{-0.98}, 0.31, 1.47^{+0.58}_{-0.43}, 3.54^{+3.31}_{-0.79})$}. 
Apart from the slope parameter $\gamma$ that is held fixed, these marginalized values are all consistent with the \citetalias{Planck13} analysis.}

Results for fitting various models to the pressure excess are displayed in Fig. \ref{fig:results}. Following the discussion in Section \ref{sec:model}, we  catalogue these results as the shock model, filament model and sub-cluster model, respectively. 
More sophisticated modelling, e.g. a combination of the sub-cluster and shock models, is beyond the scope of the current \plk analysis as the available degrees of freedom become close to zero. For each of our individual models, we compute the probability to exceed (PTE, or the $p$-value) to quantify the goodness of the fit.

\smallskip

\begin{figure}
\centering
\includegraphics[width=1.0\columnwidth]{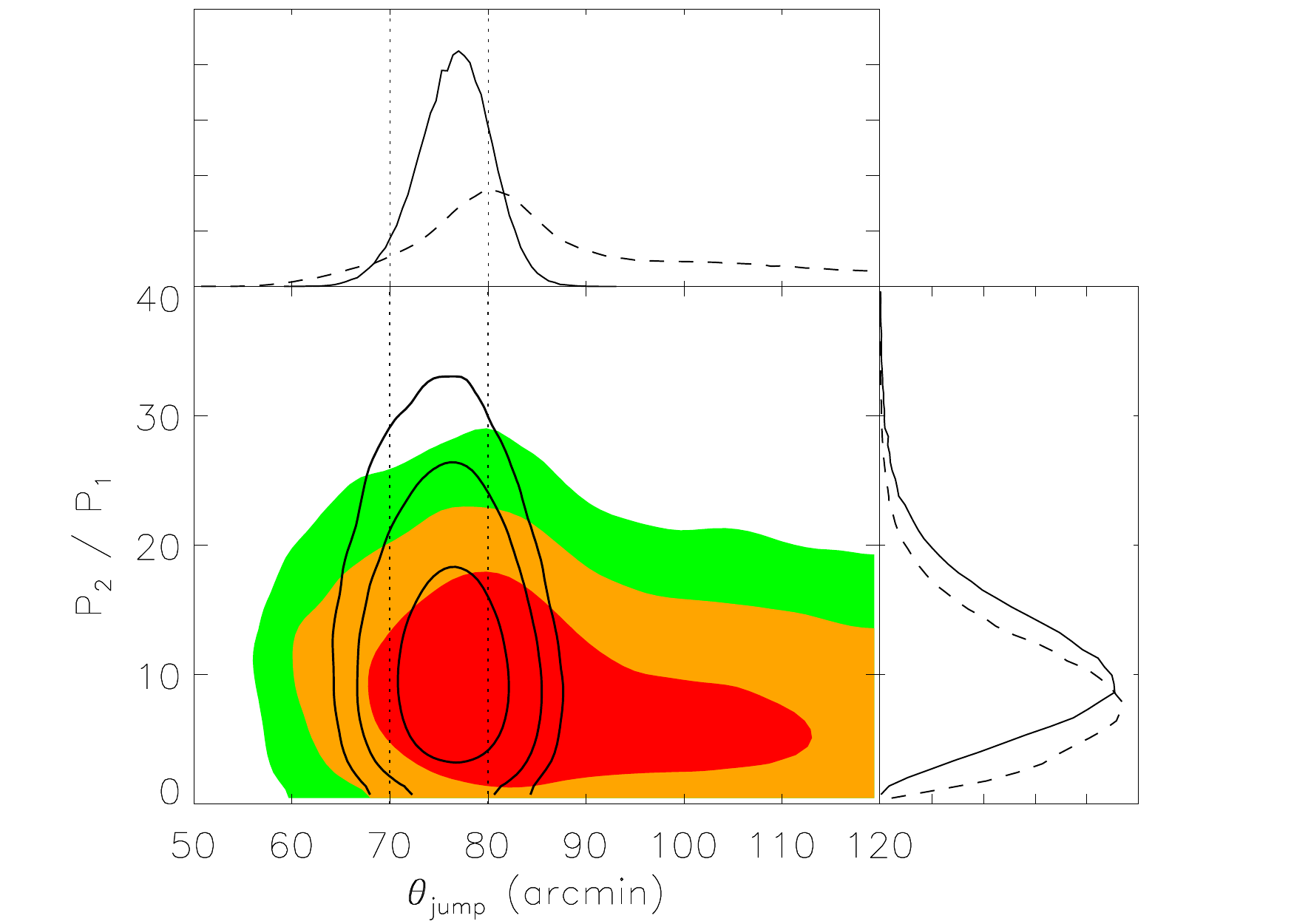}
\caption{Joint and marginalized likelihood distributions for the shock position ($\theta_{\mathrm{jump}}$) and pressure jump ratio ($P_2/P_1$) from the SZE data, with (dashed lines) and without (solid lines) putting a radio prior on the shock location. The contours indicate the 68.3, 95.5 and 99.7 per cent credibility intervals. The dotted lines indicate the FWHM of the prior on the position of the shock as derived from our radio data.}
\label{fig:likelihoods}
\end{figure}

\noindent {\it Shock model:} here, we model the pressure inside the S-W wedge as the sum of the cluster's GNFW pressure profile and a high-pressure shock tube that terminates at a given position within the fitting region. 
In a first attempt, the location of the pressure jump is treated as a free parameter,   
{\fincomm yielding a value for the shock front at $79^{+10}_{-9} \, {\rm arcmin}$ from the cluster centre. This is fully consistent with the relic position derived from radio data (75 arcmin from the cluster centre). The pressure ratio of the GNFW-model to the overlying tube corresponds to $7.3^{+5.3}_{-4.5}$ at the position of the shock front. The marginalized distributions are shown in Fig. \ref{fig:likelihoods}.
From the pressure ratio we infer the Mach number of the shock by using the Rankine-Hugoniot condition: $P_2/P_1 = ( 2 \gamma {\cal M}^2 - \gamma +1 )/(\gamma + 1)$, where $\gamma = 5/3$ is the adiabatic 
index. We obtain ${\cal M} = 2.7^{+0.8}_{-0.9}$ and a PTE $0.3$, suggesting a good fit. }

We can further constrain the shock position by multiplying our likelihood function with a Gaussian prior centred at the position of the radio relic and a FWHM of $10 \, {\rm arcmin}$ . This prior has practically no effect on the pressure ratio (or the Mach number) since once the shock width is fixed, the shock location and the compression ratio are nearly uncorrelated. {\fincomm The final result is a marginalized pressure ratio of $8.8^{+6.1}_{-3.4}$, or a Mach number ${\cal M} = 2.9^{+0.8}_{-0.6}$, with PTE $0.3$.}

{\comm We identified the pressure excess observed at the lower neighbouring wedge, near the southern tip of the relic}, with a background optical cluster at $z=0.45$ (WHL J125525.6$+$265648, Wen, Han \& Liu \citeyear{Wen09}; see Fig. \ref{fig:allcones}). Since this source is unresolved, we subtract it by modelling it with the \plk beam, which brings the pressure jump location without any positional prior more in agreement with the radio relic (at $76^{+3}_{-4} \, {\rm arcmin}$). Since this source subtraction has minimal effect on the actual pressure ratio or the Mach number, but can lead to additional systematic errors, we quote results in this Letter without any subtraction of background galaxy clusters.

\newpage

\noindent {\it Filament model:} if we let the power-law pressure excess continue without a break over the radio relic position, the PTE value for the fit drops to $0.10$. Even though this is statistically less favourable than the above shock model, a long filament with a continuous pressure gradient cannot be ruled out from \plk data. However, additional considerations based on published X-ray measurements disfavour this model. \citetalias{Akamatsu13} determined the density ratio between gas at the position of the relic and beyond it through X-ray surface brightness measurements with {\it Suzaku}. By using the pressure ratio predicted by our filament model and combining it with the observed {\comm density ratio of $\rho_2/\rho_1=2.3$ from the {\it Suzaku} data}, we find a temperature ratio of $T_1/T_2 = 1.4 \pm 0.2$, i.e. if the filament model holds true, the temperature is expected to {\it increase} beyond the relic. Such a temperature increase is incompatible with the measurements (or upper limits) from X-
ray data.

\smallskip

\noindent {\it Sub-cluster model:} to test if the infalling group NGC 4839, which lies at the {inner boundary of the relic-wedge} (Fig. \ref{fig:maps}), can be responsible for the observed pressure excess, we introduce a second GNFW component instead of a pressure tube (Fig. \ref{fig:results}, right). We put the centre of this second component at \hbox{${\rm (RA, Dec)}=(12^{\rm h}56^{\rm m}59^{\rm s}, \ +27^{\circ}29'27'')$} and fix all the shape parameters of its pressure profile to the ``universal'' shape given by Arnaud et al. \citeyearpar{Arnaud10}. The pressure amplitude of the group is fixed to 10\% of the mass of the Coma cluster, which is the upper limit from {\comm dynamical} observations (Colless \& Dunn \citeyear{Colless96}). This leads to a PTE of only 0.005, implying that NGC 4839 with 10\% of the Coma mass is insufficient to produce the observed pressure excess. If we let the mass (i.e. the pressure amplitude) of the group to float, a much better fit ($\mathrm{PTE}=0.60$) is obtained if the 
group mass is $0.28 \pm 0.03$ times the mass 
of the 
Coma 
cluster. 
Such a high mass is in conflict with the observed optical, X-ray (Neumann et al. \citeyear{Neumann01}) or Faraday rotation (Bonafede et al. \citeyear{Bonafede13}) measurements. Another unphysical assumption will be that the group retains its ``universal'' pressure shape in all directions despite the inevitable ram-pressure stripping during its infall.   

\bigskip

\vspace*{-4mm}
\section{Discussion}
\label{sec:concl}

{\comm We present the first attempt to measure a cluster merger shock at a radio relic position with the SZE.} 
Of the three models tested, a tube of high-pressure gas with a power-law pressure profile and a pressure discontinuity at the relic position provides the best-fit to the SZE data. The best fit pressure ratio suggests a shock with Mach number ${\cal M} \sim 3$, {\comm as expected for radio emitting merger shocks near cluster virial radii (e.g. Hong et al. \citeyear{Hong14})}. Our results are tied to the assumption of the pressure tube geometry, which we base on the new $2.3 \, {\rm GHz}$ radio measurement of the Coma relic. The radio data fixes the width of the propagating shock front (or the high pressure filament) at roughly 730 kpc in the plane of the sky. {\comm Disc-shaped shock fronts with a cylindrical wake} can be seen in hydrodynamical simulations of cluster mergers (e.g. Vazza et al. \citeyear{Vazza12}) and provide a simple yet realistic model for deprojecting the \plk $y$-map. 

At a significance level of $2.3 \sigma$, our pressure discontinuity measurement is not yet definitive, although it provides a useful reference for future works and comparing with other data sets. 
{\fincomm The derived Mach number is consistent within $2 \, \sigma$ with the estimates from {\it XMM-Newton} (${\cal M} = 1.9^{+0.16}_{-0.40}$, Ogrean \& Br{\"u}ggen \citeyear{Ogrean13}) and {\it Suzaku} (${\cal M} = 2.2 \pm 0.5$, \citetalias{Akamatsu13}) observations.} The current SZ-based errors are larger than the X-ray spectroscopic measurements, although different X-ray studies vary widely on their treatment of the background systematics and the quoted X-ray Mach numbers do not necessarily reflect on those systematics (e.g. Simionescu et al. \citeyear{Simionescu13} failed to derive a meaningful temperature measurement beyond the relic position using the same {\it Suzaku} data). 

Radio spectral index measurements are an alternative approach to derive the shock Mach numbers (e.g. Blandford \& Eichler \citeyear{Blandford87}). 
Taking into account the effects of particle ageing, the observed spectral index and shock Mach number relate as \hbox{${\cal M}^2 = (\alpha_{\rm obs} + 1)/(\alpha_{\rm obs} - 1)$}. We adopt the radio measurements of the Coma relic from Thierbach, Klein \& Wielebinski \citeyearpar{Thierbach03}, who found \hbox{$\alpha_{\rm obs} = 1.18 \pm 0.02$}. This implies a radio Mach number \hbox{${\cal M}=3.5\pm0.2$}, ignoring possible systematics from low-resolution single dish radio data and multi-frequency flux comparison. {\fincomm The SZ-derived value for the shock Mach number thus seems to lie in between the radio and X-ray measurements,  
but the current \plk data are unsuitable for a more precise comparison.}

Our SZE analysis (with radio priors) supports the idea of an outwardly moving shock front, likely caused by the first infall of the group NGC 4839. The inward moving shock is lost in the high-density central region of the Coma cluster (e.g. Markevitch \& Vikhlinin \citeyear{Markevitch07}). 
However, the causal connection between the infalling group and the shock front detected at the radio relic remains speculative and we do not claim to know the origin of the shock that is likely causing the radio emission in the Coma relic.  

{\comm Presence of a sharp SZE feature will have some impact on the high-frequency radio measurements of cluster radio relics (up to $15$ or $20 \, {\rm GHz}$), that aim to detect a spectral steepening to constrain the particle injection and energy loss models (e.g. Stroe et al. \citeyear{Stroe14}). The SZE decrement signal boosted by the pressure jump will 
produce a flux offset for the radio synchrotron emission. This can potentially lead to a false determination of the spectral steepening in the radio spectrum.}  

\newpage

\vspace*{-4mm}
\section*{Acknowledgements}

We gratefully acknowledge Lawrence Rudnick and Shea Brown for providing the WSRT 350 MHz radio image, and Franco Vazza, Annalisa Bonafede and Florian Pacaud for discussions. We acknowledge financial support by the DFG: UK and MT in the framework of the Forschergruppe 1254, KB and FB through the TRR 33 and FB also through the CRC 956.

\label{lastpage}
\end{document}